\documentclass[iop]{emulateapj}

\journalinfo{Submitted to ApJ}
\submitted{Received 2014 February 27}

\shorttitle{Tests Of In-Situ Multiplanet Formation}
\shortauthors{Schlaufman}

\begin{document}

\title{Tests of In-Situ Formation Scenarios for Compact Multiplanet Systems}

\author{Kevin C.\ Schlaufman\altaffilmark{1}}
\affil{Kavli Institute for Astrophysics and Space Research,
Massachusetts Institute of Technology, Cambridge, MA 02139, USA}
\email{kschlauf@mit.edu}

\altaffiltext{1}{Kavli Fellow}

\begin{abstract}

\noindent
{\it Kepler} has identified over 600 multiplanet systems, many of which
have several planets with orbital distances smaller than that of Mercury
-- quite different from the Solar System.  Because these systems may be
difficult to explain in the paradigm of core accretion and disk migration,
it has been suggested that they formed in situ within protoplanetary
disks with high solid surface densities.  The strong connection between
giant planet occurrence and stellar metallicity is thought to be linked
to enhanced solid surface densities in disks around metal-rich stars,
so the presence of a giant planet can be a detectable sign of planet
formation in a high solid surface density disk.  I formulate quantitative
predictions for the frequency of long-period giant planets in these in
situ models of planet formation by translating the proposed increase in
disk mass into an equivalent metallicity enhancement.  I rederive the
scaling of giant planet occurrence with metallicity as $P_{\mathrm{gp}}
= 0.05_{-0.02}^{+0.02} \times 10^{(2.1 \pm 0.4) [\mathrm{M/H}]} =
0.08_{-0.03}^{+0.02} \times 10^{(2.3 \pm 0.4) [\mathrm{Fe/H}]}$ and
show that there is significant tension between the frequency of giant
planets suggested by the minimum mass extrasolar nebula scenario and
the observational upper limits.  This fact suggests that high-mass disks
alone cannot explain the observed properties of the close-in {\it Kepler}
multiplanet systems and that migration is still a necessary contributor to
their formation.  More speculatively, I combine the metallicity scaling of
giant planet occurrence with recently published small planet occurrence
rates to estimate the number of Solar System analogs in the Galaxy.
I find that in the Milky Way there are perhaps $4 \times 10^{6}$ true
Solar System analogs with an FGK star hosting both a terrestrial planet
in the habitable zone and a long-period giant planet companion.

\end{abstract}


\section{Introduction}

{\it Kepler} has discovered many multiplanet systems with several planets
with orbital periods $P < 50$ days.\footnote{See for example
   \citet{bor11a,bor11b}, \citet{bat13}, and \citet{bur14}.}
Indeed, 40\% of solar-type stars in the {\it Kepler} field have at least
one planet with $P < 50$ days \citep[e.g.,][]{fre13}.  Even though these
systems differ from the Solar System, their apparent ubiquity suggests
that they may represent a frequent outcome of planet formation.

In the traditional minimum mass solar nebula (MMSN) scenario, there is
probably insufficient solid material in protoplanetary disks to form
the {\it Kepler} multiplanet systems where they are observed today
\citep{wei77,hay81}.  Instead, formation further out in the parent
protoplanetary disk combined with subsequent inward migration has been
suggested as one possible formation channel for this class of system
\citep[e.g.,][]{ali06}.  The apparent excess of planets just outside
of mean-motion resonances may also support the formation then inward
migration scenario \citep[e.g.,][]{lis11,fab12}.  However, the rate
and even direction of migration is known to sensitively depend on the
unknown thermodynamic state of the disk \citep[e.g.,][]{paa10,kle12}.

\begin{figure*}
\plotone{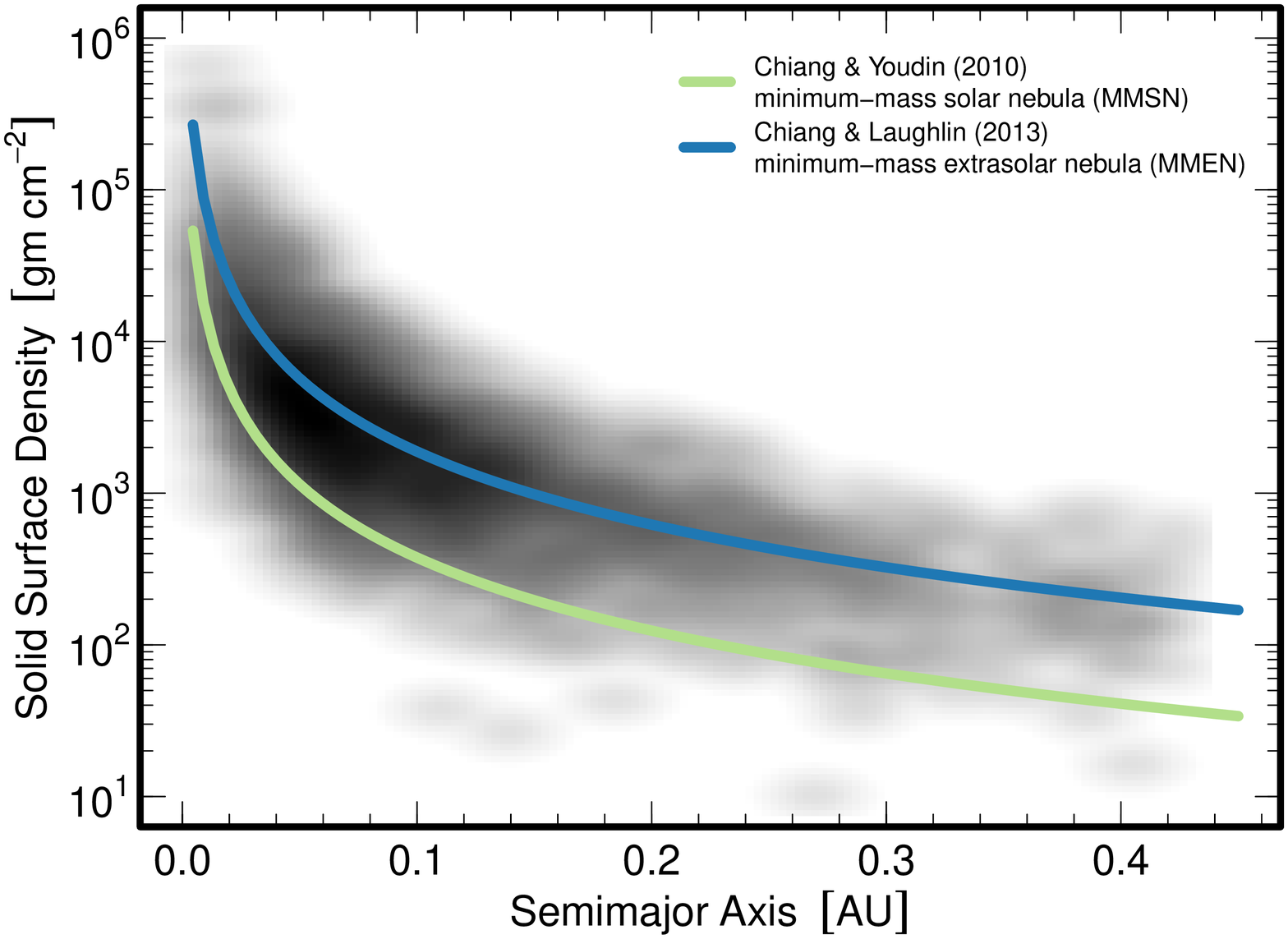}
\caption{Solid surface density profile $\Sigma_{\textrm{solid}}$ of the
\citet{chi13} minimum-mass extrasolar nebula (MMEN).  The background
shading indicates the density of points in the semimajor axis--solid
surface density diagram for \textit{Kepler} objects of interest (KOIs).
I define the solid surface density of a KOI as $\Sigma_{\textrm{solid}}
\equiv M_{p}/(2 \pi a_{p}^2)$.  I compute the mass of each KOI by assuming
$M_{p} = R_{p}^{2.06}$ \citep[e.g.,][]{lis11} and each semimajor axis
$a_{p}$ using the observed period $P$ and assuming a $1~M_{\odot}$
host star.  The green curve shows the standard \cite{hay81} minimum-mass
solar nebula (MMSN) as parametrized by \cite{chi10}, while the blue
curve shoes the fiducial MMEN of \citet{chi13}.  There is insufficient
solid material in the MMSN nebula to form the observed KOIs in situ,
so migration of solids is required to explain the KOIs.  On the other
hand, there is sufficient solid material present in the MMEN scenario
to form the KOIs in situ.\label{fig01}}
\end{figure*}

Alternative models of in situ formation in disks with solid surface
densities enhanced beyond the MMSN expectation have also been suggested
to explain the ubiquity of close-in multiple systems.  In the minimum
mass extrasolar nebula (MMEN) scenario of \citet{chi13}, the {\it Kepler}
multiplanet systems formed in protoplanetary disks that were about six
times more massive than those envisioned in the MMSN scenario.  In that
picture, the more massive MMEN disks describe the typical protoplanetary
disk in the Galaxy, while the less massive MMSN disk is the outlier.
I illustrate the MMEN in Figure~\ref{fig01}.  On the other hand,
\citet{han12} invoke the rapid inward migration of planetesimals into
the inner regions of the disk.  The enhanced solid surface density of
the inner disk then naturally leads to the in situ formation of planetary
systems closely resembling those observed by {\it Kepler} \citep{han13}.

Both models of in situ planet formation described above provide useful,
fresh looks at planet formation.  As I will show, both models are
also amenable to quantitative tests.  At face value, both models make
qualitative predictions for the formation of long-period giant planets.
All else being equal, a disk with higher solid surface density in the
giant planet forming region has a better chance of forming a giant planet
than does a disk with lower solid surface density in the giant planet
forming region \citep[e.g.,][]{lis09}.  Consequently, \citet{chi13}
suggested that the enhanced solid surface density in the MMEN scenario
should lead to the efficient formation of giant planets outside of 1 AU.
In contrast, the concentration of a significant amount of a disk's solid
material in the inner disk as suggested by \citet{han12} should lead to
inefficient formation of giant planets outside of 1 AU.  Unfortunately,
it is not currently possible to directly assess with either the transit
or radial velocity technique the frequency of long-period giant planets
in the observed {\it Kepler} multiple systems themselves.

However, it is possible to indirectly infer the frequency of long-period
giant planets in at least two ways.  First, the frequency can be
characterized by proxy, a technique in which objects in the solar
neighborhood that can be studied in detail stand in for the more distant
{\it Kepler} objects.  In this case, the close-in multiple systems
of low-mass planets discovered in the solar neighborhood with the
radial velocity technique are a proxy for the more distant {\it Kepler}
multiple systems.  One can use the published completeness limits of the
radial velocity surveys to establish upper limits on the frequency of
long-period giant planets in those systems, then compare that upper limit
to quantitative predictions of the in situ models of planet formation.
Second, it well known that giant planet host stars are preferentially
metal enriched \citep[e.g.,][]{san04,fis05}.  Therefore, the absence
of a metallicity effect in solar-type hosts of {\it Kepler} multiple
systems can be used to determine an upper limit on the frequency of
giant planets in these systems.

In this paper, I compare statistical upper limits on the frequency of
long-period giant planets in the {\it Kepler} multiple systems with
quantitative predictions of the in situ models of planet formation.
I find that there is significant tension between the derived upper limits
and the expectation from the MMEN scenario, though current samples are
not large enough to constrain the \citet{han12} scenario.  I describe
my sample selection in Section 2, I detail my statistical analyses in
Section 3, I discuss the results and implications in Section 4, and I
summarize my findings in Section 5.

\section{Sample Definition}

I select \textit{Kepler} multiple planet systems having at least two
planets with $R_{\mathrm{P}} < 5~R_{\oplus}$ from the {\it Kepler}
CasJobs database\footnote{http://mastweb.stsci.edu/kplrcasjobs/guide.aspx}
using the query given in the Appendix.  I focus exclusively on exoplanet
systems orbiting solar-type stars, because the MMEN scenario is scaled
from the solar nebula and because there is both a theoretical expectation
and observational evidence that the planet formation process changes for
low-mass stars \citep[e.g.,][]{lau04,ida05,end06,but06,bon13}.  For that
reason, I select systems orbiting solar-type stars with $0.22 < J-H <
0.62$, $0.00 < H-K < 0.10$, and $0.22 < J-K < 0.72$ roughly corresponding
to spectral types in the range F5--K5 \citep{cov07}.  From here, I refer
to the systems in this sample as the \textit{Kepler} multiplanet systems.

\begin{figure*}
\plottwo{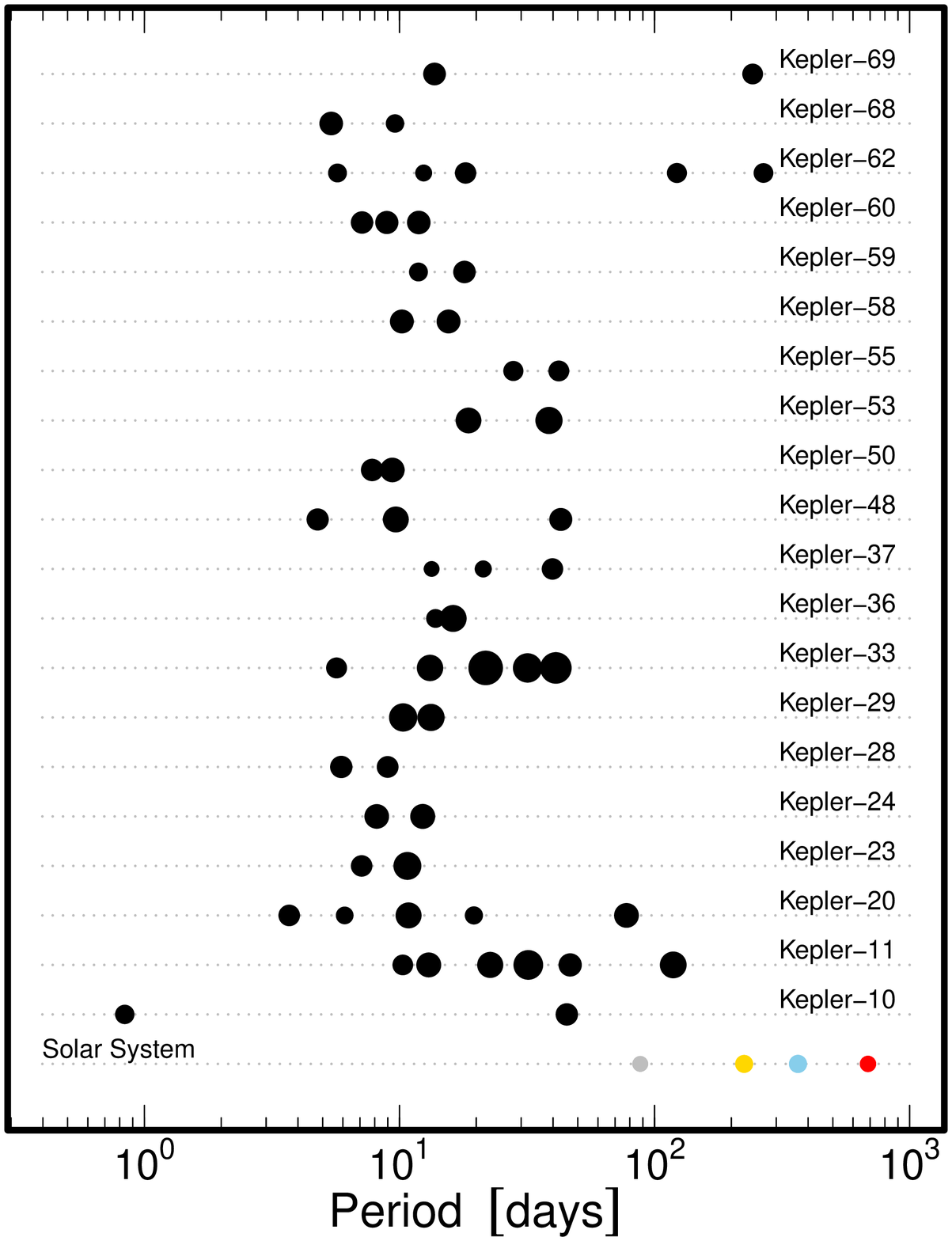}{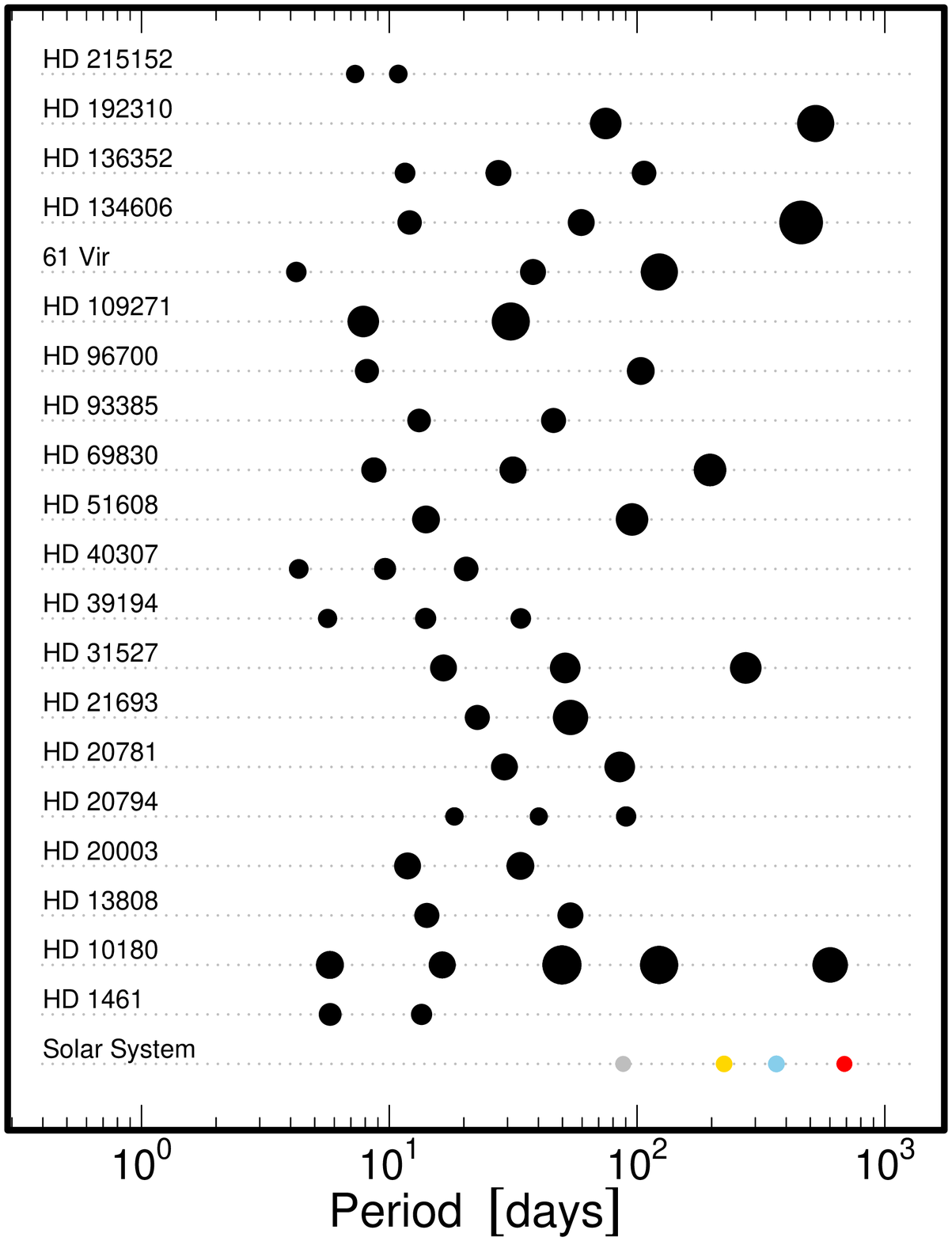}
\caption{Multiplanet system architectures.  \textit{Left}: Validated
\textit{Kepler} multiplanet systems.  The size of each planet is
proportional to its estimated radius.  The terrestrial planets in the
Solar System are included for scale.  \textit{Right}: Systems of multiple
Neptune-mass planets discovered with the radial velocity technique (RV
multiplanet systems).  The size of each planet is proportional to its
estimated minimum mass.  The terrestrial planets in the Solar System
are included for scale.  The architectures of the \textit{Kepler} and
RV multiplanet systems are similar, indicating that the properties of RV
multiplanet systems can reasonably be used as proxies for the properties
of the \textit{Kepler} multiplanet systems.\label{fig02}}
\end{figure*}

Likewise, I select systems of exoplanets discovered with the radial
velocity technique that have at least two planets with $m\sin{i} \leq
0.1\,M_{\textrm{Jup}} = 31.8\,M_{\oplus}$ from both exoplanets.eu
and exoplanets.org \citep{schn11,wri11}.  For each planet host
star, I obtain \textit{Hipparcos} parallaxes and $B-V$ colors from
\citet{van07} and apparent Tycho-2 $V$-band magnitudes from \citet{hog00}.
I transform Tycho-2 $B_{T}$ and $V_{T}$ magnitudes into approximate
Johnson--Cousins $V$-band magnitudes using the relation $V = V_{T}
- 0.090\left(B_{T}-V_{T}\right)$.  I then select exoplanet systems
orbiting solar-type stars with $0.44 < B-V < 1.15$ and $3.5 < M_{V}
< 7.4$, roughly corresponding to spectral types in the range F5--K5
\citep{bin98}.  I give the exoplanet systems and host star properties
that result from this selection in Tables~\ref{tbl-1} and~\ref{tbl-2}.
From here, I refer to the systems in this sample as the RV multiplanet
systems.  I plot both samples in Figure~\ref{fig02}.

\section{Analysis}

\subsection{Giant Planet Formation}

It is well established that giant planets occur more frequently around
metal-rich stars \citep[e.g.,][]{san04,fis05}.  The high metallicity
of a star is interpreted as evidence that its parent disk was enriched
in dust.  A dust-enriched protoplanetary disk is thought to be more
likely to form the $\approx\!\!10~M_{\oplus}$ core necessary for
giant planet formation in the few million years available before the
disk disappears.  Giant planets are also less frequently found around
M dwarfs than solar-type stars \citep[e.g.,][]{end06,but06,bon13}.
M dwarfs presumably formed from lower-mass disks than solar-type stars,
and the reduced amount of dust present in a low-mass disk is thought to
make it more difficult to assemble a core.  Recent observational evidence
that disk mass scales approximately linearly with host star mass below
$1~M_{\odot}$ supports this view \citep{and13}.

In short, giant planet formation efficiency is proportional to the solid
surface density in a protoplanetary disk $\Sigma_{\mathrm{solid}}$, where

\begin{eqnarray}
\Sigma_{\mathrm{solid}} \propto f_{\mathrm{dust}} M_{\mathrm{disk}}
                        \propto Z M_{\mathrm{disk}}.
\end{eqnarray}

\noindent
The first proportionality is true by construction.  The second is true
because observations of the LMC and Milky Way show that the dust-to-gas
fraction $f_{\mathrm{dust}}$ scales approximately linearly with metal mass
fraction $Z$ \citep{gor03}.  Consequently, the six times increase in disk
mass in the MMEN scenario relative to the MMSN scenario should have an
equivalent effect on giant planet formation efficiency as increasing $Z$
by a factor of six, or increasing [M/H] by 0.78 dex.

The pioneering work of \citet{fis05} showed that the probability of
giant planet occurrence $P_{\mathrm{gp}}$ is

\begin{eqnarray}
P_{\mathrm{gp}} = 0.03 \times 10^{2.0\mathrm{[Fe/H]}}.
\end{eqnarray}

\noindent
They argued that the dependence of $P_{\mathrm{gp}}$ on the square of
the number of iron atoms present was the natural expectation from the
collisional agglomeration of dust grains into planetesimals.  However,
their result lacked an uncertainty estimate and may have been affected by
the need to bin their data.  Moreover, many new, longer-period giant
planets have been discovered in the interim, so a new calculation
is timely.  For those reasons, I use logistic regression to rederive
the scaling of giant planet occurrence with metallicity, as logistic
regression both avoids the need to bin the data and naturally produces
an error estimate on the scaling \citep[e.g.,][]{cha00}.

As input, I use a sample of 1111 FGK field and planet host stars from
the HARPS GTO planet search program from \citet{adi12b}\footnote{I
   arrive at quantitatively similar results if I instead use the
   \citet{val05} SPOCS sample.}.  Stellar parameters and abundances for
   each star in the catalog have been homogeneously derived using the
techniques presented in \citet{sou08,sou11a,sou11b}.
I cross-match all of the stars with the {\it
Hipparcos}, Tycho-2, and exoplanets.org catalogs using
\texttt{TOPCAT}\footnote{http://www.star.bristol.ac.uk/\textasciitilde
mbt/topcat/} \citep{tay05}.  I retain only those stars with distance $d <
50$ pc from the Sun, parallaxes precise to 20\%, $0.44 < B-V < 1.15$,
and $3.5 < M_{V} < 7.4$.  Finally, I code as giant planet host stars
those stars with planets with radial velocity semiampitude $K > 20$
m s$^{-1}$ and $P < 4$ years, because the completeness of the survey is
very high in that range of parameter space.  The end result is a sample
of 620 solar-type stars, 44 of which host at least one giant planet.

The probability of giant planet occurrence must be in the interval $0
\leq P_{\mathrm{gp}} \leq 1$.  Linear regression is unsuitable for the
prediction of probabilities, because the linear regression equation $y
= \beta_{0} + \sum \beta_{i} x_{i}$ is not bounded between 0 and 1.
Instead, logistic regression makes use of the logistic function to
predict the probability

\begin{eqnarray}
P(Y) = \frac{e^{\beta_{0} + \sum \beta_{i} x_{i}}}{
             1+e^{\beta_{0} + \sum \beta_{i} x_{i}}}.
\end{eqnarray}

\begin{figure}
\plotone{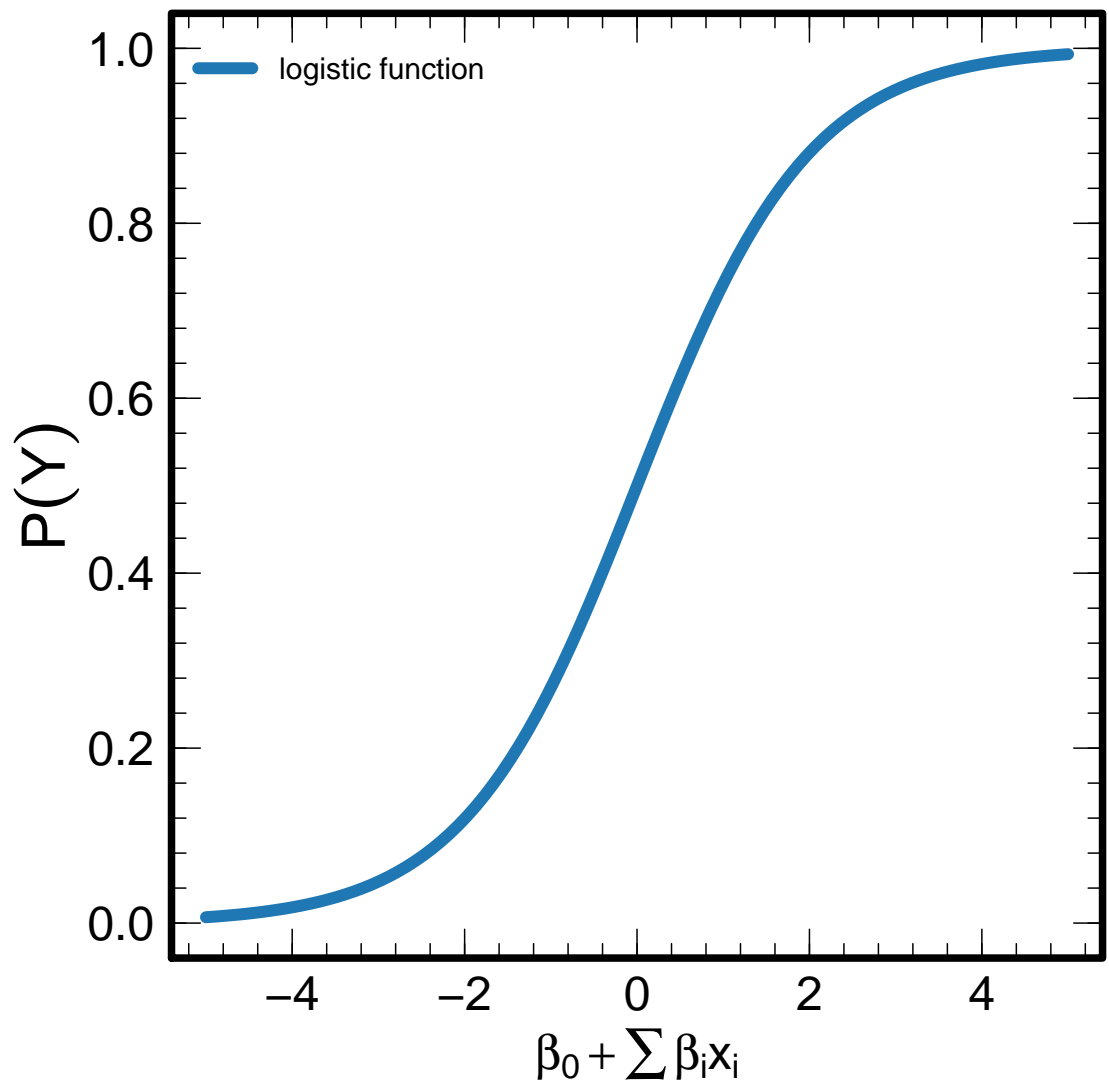}
\caption{The logistic function.\label{fig03}}
\end{figure}

\noindent
I plot the logistic function in Figure~\ref{fig03}.  It takes any real
value $x$ and maps it into the interval $0 \leq y \leq 1$, satisfying
the requirement that the probability of an event be between 0 and 1.

The logistic function is nonlinear.  As a result, logistic regression
works with the natural logarithm of the odds ratio

\begin{eqnarray}
\log{\left[\frac{P(Y)}{1-P(Y)}\right]} = \beta_{0} + \sum \beta_{i} x_{i},
\end{eqnarray}

\noindent
or the logit function, which is linear in the coefficients $\beta_{i}$.
The coefficients $\beta_{i}$ can then be fit numerically.  For $i>0$,
the interpretation of the coefficient $\beta_{i}$ is that a one unit
change in $x_{i}$ changes the log odds ratio of the probability of
the event by a factor of $\beta_{i}$, or the probability of the event
itself by $e^{\beta_{i}}$.  The coefficient $\beta_{0}$ has no meaningful
interpretation.

In this context, I calculate the response of $P_{\mathrm{gp}}$ to only
one predictor: either $x_{1}$ = [M/H] or $x_{1}$ = [Fe/H].  I use the
\texttt{glm} function in \texttt{R}\footnote{http://www.R-project.org/}
to compute a logistic regression model \citep{r13}.  I build models
predicting the effect of both [M/H] and [Fe/H] on the probability of
giant planet occurrence.  I compute [M/H] assuming the solar abundances
from \citet{asp05}.  I find that

\begin{eqnarray}
P_{\mathrm{gp}}([\mathrm{M/H}]) & \propto & 10^{(2.1 \pm 0.4) \Delta [\mathrm{M/H}]}, \\
P_{\mathrm{gp}}([\mathrm{Fe/H}])& \propto & 10^{(2.3 \pm 0.4) \Delta [\mathrm{Fe/H}]}.
\end{eqnarray}

\noindent
To determine the absolute probability of giant planet occurrence at
[M/H] = [Fe/H] = 0, I calculate the fraction of stars with giant
planets in the sample in the ranges $-0.05 < [\mathrm{M/H}] < 0.05$
and $-0.05 < [\mathrm{Fe/H}] < 0.05$.  I use bootstrap resampling to
determine confidence intervals, and I find that near $[\mathrm{M/H}] =
[\mathrm{Fe/H}] = 0$

\begin{eqnarray}
P_{\mathrm{gp}}([\mathrm{M/H}]) & = & 0.05_{-0.02}^{+0.02} \times 10^{(2.1 \pm 0.4) [\mathrm{M/H}]}, \\
P_{\mathrm{gp}}([\mathrm{Fe/H}])& = & 0.08_{-0.03}^{+0.02} \times 10^{(2.3 \pm 0.4) [\mathrm{Fe/H}]}.
\end{eqnarray}

\begin{figure*}
\plottwo{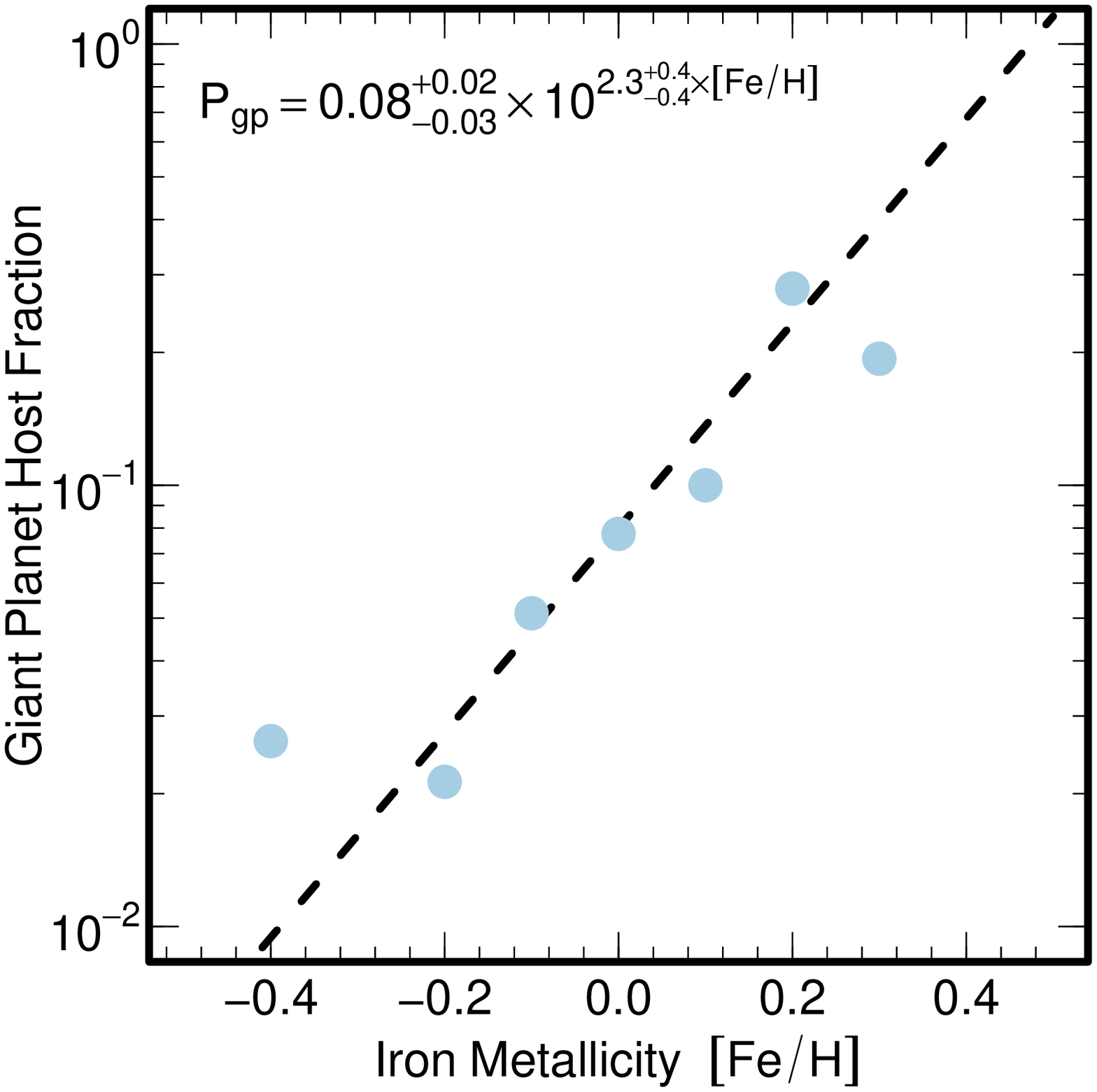}{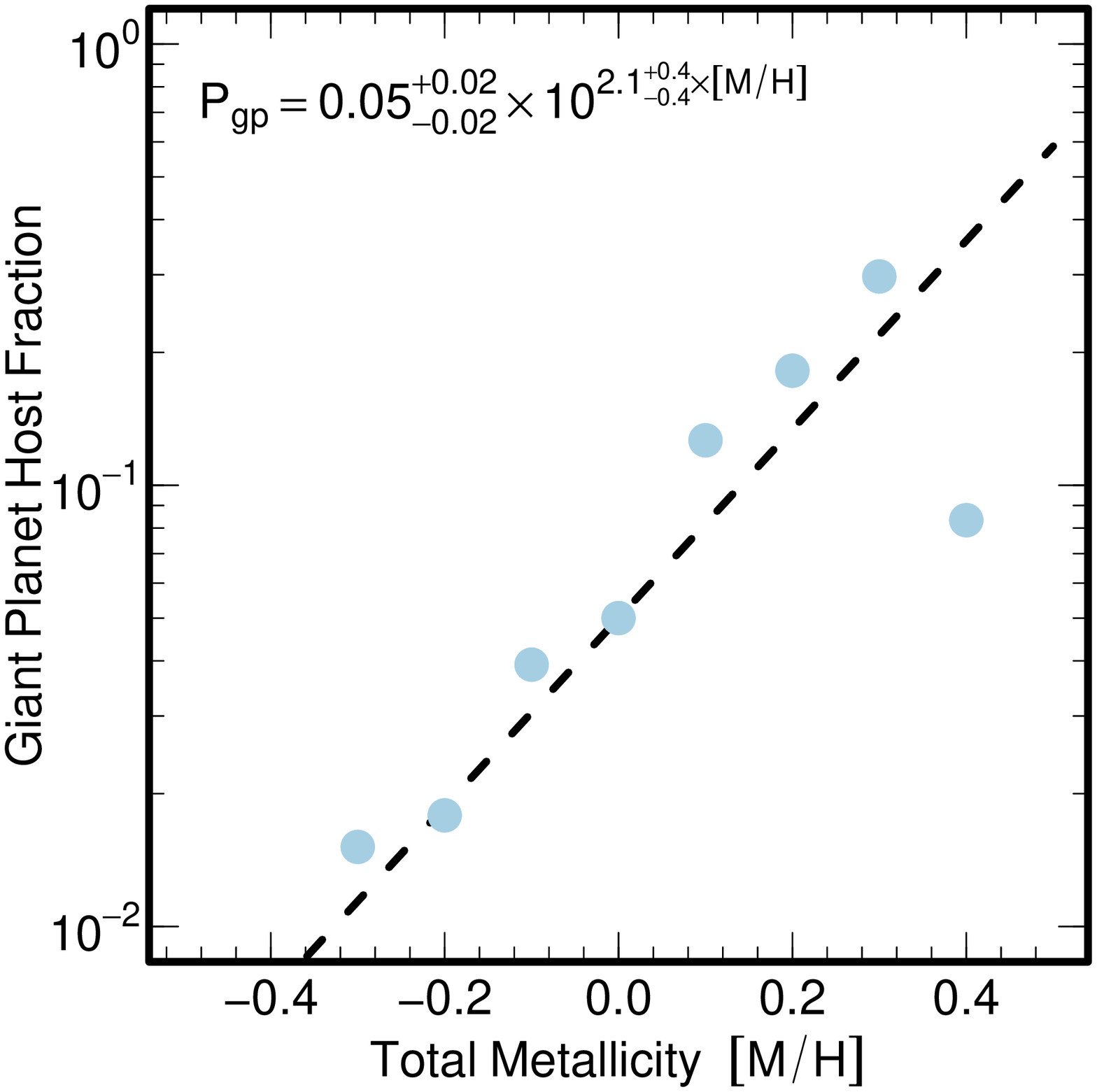}
\caption{The effect of host star composition on giant planet occurrence.
I use logistic regression to estimate the effect of composition on giant
planet occurrence, so the arbitrary binning reflected in the plots does
not affect the derived scaling.  \textit{Left}: The effect of iron
metallicity on giant planet occurrence.  There is a hint that giant
planet occurrence levels-out in the low-metallicity tail of the thin
disk metallicity distribution \citep[e.g.,][]{san04}.  \textit{Right}:
The effect of total metallicity on giant planet occurrence.  The hint
of a plateau at low metallicity disappears when considering total
metallicity.  This occurs because non-solar abundance patterns (e.g.,
$\alpha$-enhancement) begin to appear in stars with [Fe/H] $\approx -0.4$
as the thick disk stellar population starts to become considerable relative
to the thin disk \citep[e.g.,][]{adi12a}.\label{fig04}}
\end{figure*}

\noindent
I plot the result in Figure~\ref{fig04}.

These results suggest that in the MMEN scenario, the probability of
giant planet occurrence in the {\it Kepler} multiplanet systems should
be $P_{\mathrm{gp}}(0.78) \approx 1$.  Given that these giant planets
would be found at $a \gtrsim 1$ AU, they would only rarely be observed
to transit and therefore be nearly invisible to the transit technique.
However, these planets would likely be detectable with the radial
velocity technique.

\subsection{Stability and Completeness Constraints}

The analysis in Section 3.1 indicates that practically every {\it
Kepler} multiplanet system formed in the MMEN scenario should have
a long-period giant planet.  However, it is currently impractical to
search for long-period giant planets in the {\it Kepler} multiplanet
systems themselves.  Instead, I use the RV multiplanet systems in
Table~\ref{tbl-2} as a proxy and use the published RV completeness
estimates for those systems to derive a constraint on the frequency of
long-period giant planets in the {\it Kepler} multiplanet systems.

The existence and stability of the {\it Kepler} multiplanet systems with
orbital periods $P<50$ days precludes the existence of giant planets
with similar orbital periods.  On the other hand, longer-period giant
planets are not prohibited by stability arguments.  A multiplanet
system is likely to be stable if its planets are separated by 10 or
more mutual Hill radii $R_{H} = \left[M_p/(3 M_{\ast})\right]^{1/3}$
\citep[e.g.,][]{cha96,smi09}.  For a system of multiple planets orbiting
a $1~M_{\odot}$ star with its outermost Neptune-mass planet $M_{1} =
17.147\,M_{\oplus}$ at an orbital period of $P_{1} = 50$ days

\begin{eqnarray}
a_{1} & = & \left(\frac{P_{1}}{365}\right)^{2/3}, \\
A & = & a_{1} + 10\left(\frac{M_{1}}{3 M_{\odot}}\right)^{1/3} a_{1},
\end{eqnarray}

\noindent
the smallest semimajor axis $a_{2}$ at which a $M_{2} =
1\,M_{\mathrm{Jup}}$ giant planet would not make the system obviously
unstable is

\begin{eqnarray}
a_{2} = A \left[1-10\left(\frac{M_{2}}{3 M_{\odot}}\right)^{1/3}\right]^{-1}.
\end{eqnarray}

\noindent
I find that $a_{2} \approx 1$ AU, or $P_{2} \approx 365$ days.  About 75\%
of the giant planets identified around Solar-type stars with the radial
velocity technique have $P \geq 365$ days \citep{cum08}, and therefore
would not render a {\it Kepler} multiplanet system obviously unstable.
For that reason, the fraction of the observed giant planet systems
permitted by stability considerations is $\eta_{\mathrm{Hill}} = 0.75$.
\citet{schm13} and \citet{cab14} recently identified KOI-351 as one
such system.

\begin{figure}
\plotone{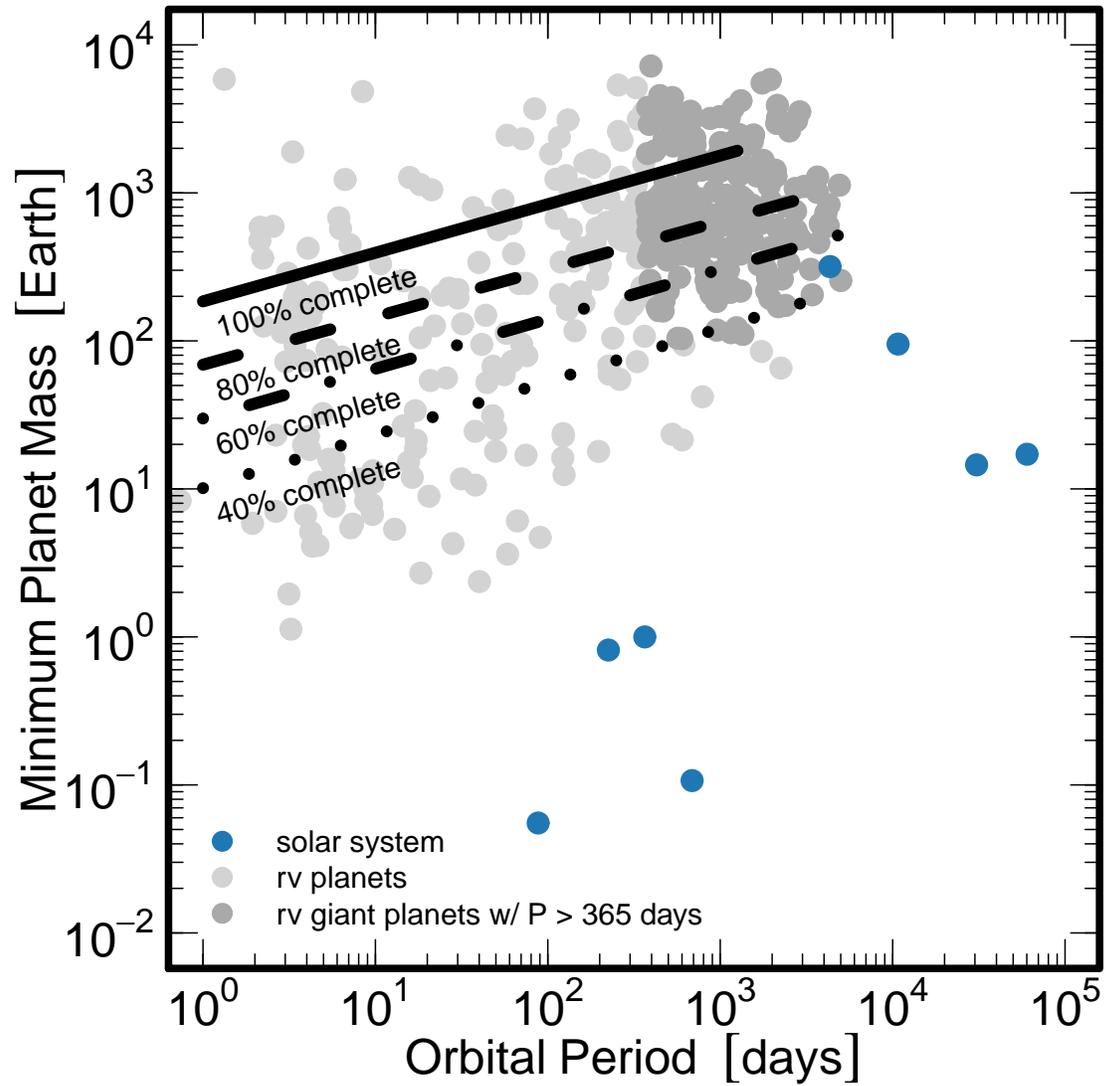}
\caption{Distribution of planets discovered with the radial velocity
technique in the semimajor axis--minimum mass plane.  I plot average
completeness curves from \citet{may11} indicating the fraction of
planetary systems above each curve that would likely have been detected
by HARPS+CORALIE.  Close-in multiplanet systems with $P < 50$
days are likely only stable if there are no giant planets in those
system with $P \lesssim 365$ days.  Therefore, I only consider giant
planets with orbital periods longer than one year in my Monte Carlo
calculations.  Fully 75\% of giant exoplanets are observed in this period
range \citep[e.g.,][]{cum08}, and I plot those planets included in my 
Monte Carlo in dark gray.  The planets in the Solar System are indicated
as blue dots.\label{fig05}}
\end{figure}

I use the completeness contours in Figure 6 of \citet{may11} to determine
the expected number of long-period giant planets that would have been
discovered around the RV multiplanet systems if every system had a
long-period giant planet with mass and period drawn from the observed
distributions of those quantities.  Since all 20 RV multiplanet systems
are in the catalog of \citet{may11}, the completeness contour given in the
paper apply for each system.  For all $p$ planets in the exoplanets.org
catalog with $m\sin{i} > 100\,M_{\oplus}$ and $P>365$ days, I check
whether each is above the (100\%,95\%,80\%,60\%,40\%) completeness
contours.  If so, I increment the expected number of detections $q$ by
(1,0.95,0.8,0.6,0.4).  If the planet is below the 40\% contour, I assume
that it would be undetectable.  I find that the fraction of giant planet
systems that would have been recovered by \citet{may11} is $q/p = \eta_{c}
= 0.71$.  I illustrate this calculation in Figure~\ref{fig05}.

\subsection{Inference}

The goal is to determine both the posterior distribution for
$P_{\mathrm{gp}}$ inferred from from the non-detection of giant planets
in the RV multiplanet systems and the posterior of $P_{\mathrm{gp}}$
expected under the MMEN scenario after taking into account stability
and completeness.  If there is tension between the two posteriors, then
the MMEN scenario may not be an accurate description of {\it Kepler}
multiplanet system formation.

Bayes' Theorem guarantees

\begin{eqnarray}
f(\theta|\vec{y}) = \frac{f(\vec{y}|\theta) f(\theta)}
                         {\int f(\vec{y}|\theta) f(\theta) d\theta},
\end{eqnarray}

\noindent
where $f(\theta|\vec{y})$ is the posterior distribution of the model
parameter $\theta$, $f(\vec{y}|\theta)$ is the likelihood of the data
$\vec{y}$ given $\theta$, and $f(\theta)$ is the prior for $\theta$.
In this case, the likelihood is the binomial likelihood that describes
the probability of a number of successes $y$ in $n$ Bernoulli trials
each with probability $\theta$ of success

\begin{eqnarray}
f(y|\theta) = \left(\begin{array}{cc} n \\ y \end{array} \right)
              \theta^{y} \left(1-\theta\right)^{n-y}.
\end{eqnarray}

\noindent
The calculation of the posterior $f(\theta|\vec{y})$ can be greatly
simplified by the selection of an appropriate prior $f(\theta)$.  In this
case, it is possible to use a conjugate prior -- a prior that guarantees
that the posterior distribution will be in the same family as the prior.
The Beta($\alpha,\beta$) distribution is a conjugate prior to the binomial
likelihood and will give a Beta posterior, where $\alpha$ and $\beta$
are the standard parameters of the Beta distribution.  I take

\begin{eqnarray}
f(\theta) & = & \mathrm{Beta}(\alpha,\beta), \\
          & = & \frac{\Gamma(\alpha+\beta)}{\Gamma(\alpha) \Gamma(\beta)}
                \theta^{\alpha-1} \left(1-\theta\right)^{\beta-1}
                I_{0 \leq \theta \leq 1},
\end{eqnarray}

\noindent
where $\Gamma(x)$ is the standard gamma function and $I_{0 \leq \theta
\leq 1}$ is the indicator function that is 1 in the interval $[0,1]$
and 0 elsewhere.

Plugging $f(y|\theta)$ and $f(\theta)$ into Bayes' Theorem shows
that the posterior distribution $f(\theta|y)$ can be written

\begin{eqnarray}
f(\theta|y) & = & \frac{\Gamma(\alpha+y+\beta+n-y)}
                       {\Gamma(\alpha+y) \Gamma(\beta + n -y)}
                  \theta^{\alpha+y-1} \left(1-\theta\right)^{\beta+n-y-1}
                  I_{0 \leq \theta \leq 1}, \\
            & = & \mathrm{Beta}(\alpha+y,\beta+n-y).
\end{eqnarray}

\noindent
In words, the posterior distribution of $\theta$ is itself a Beta
function.  The hyperparameters $\alpha$ and $\beta$ of the prior can
be thought of as encoding a certain amount of prior information in the
form of pseudo-observations.  Specifically, $\alpha-1$ is the number of
success and $\beta-1$ is the number of failures imagined to have already
been observed and therefore included as prior information on $\theta$.
Taking any $\alpha = \beta= i$ where $i \geq 1$ could be thought of as
an uninformative prior in the sense that the probability of success and
failure in the prior distribution are equally likely.  However, if $i$
is large then there is imagined to be a lot of prior information and the
posterior distribution will mostly reflect the prior when $n \leq i$.
On the other hand, if $n \gg i$, then the posterior will be dominated
by the data.  For that reason, I take $\alpha=\beta=1$.

In the exoplanet context, $\theta$ is the unknown probability of
giant planet occurrence $P_{\mathrm{gp}}$, $n=20$ is the number of
RV multiplanet systems in Table~\ref{tbl-2}, and $y$ is the number
of detected long-period giant planets.  No giant planets have been
detected in the RV multiplanet systems, so $y=0$.  After accounting for
completeness and stability, the equivalent number of systems searched
at 100\% completeness would be $n' = \eta_{c} \, \eta_{\mathrm{Hill}} \,
n \approx (0.71)(0.75)(20) \approx 10.6$.  The posterior distribution of
$P_{\mathrm{gp}}$ inferred from the non-detection of long-period giant
planets in the RV multiplanet systems is

\begin{eqnarray}
P_{\mathrm{gp,obs}} & = & \mathrm{Beta}(\alpha+y,\beta+n'-y), \\
                    & = & \mathrm{Beta}(1+0,1+\eta_{c} \eta_{\mathrm{Hill}} n - 0), \\
                    & = & \mathrm{Beta}(1,11.6).
\end{eqnarray}

\noindent
In the MMEN scenario, every RV multiplanet system should have a
long-period giant planet.  Consequently, the expected number of
successful planet discoveries after examining $n$ systems should be $y'
= \eta_{c} \, \eta_{\mathrm{Hill}} \, n$.  The posterior distribution
of $P_{\mathrm{gp}}$ in the MMEN scenario should be

\begin{eqnarray}
P_{\mathrm{gp,MMEN}} & = & \mathrm{Beta}(\alpha+y',\beta+n-y'), \\
                    & = & \mathrm{Beta}(1+\eta_{c} \eta_{\mathrm{Hill}} n,1 + n - \eta_{c} \eta_{\mathrm{Hill}} n), \\
                    & = & \mathrm{Beta}(11.6,10.4).
\end{eqnarray}

\begin{figure}
\plotone{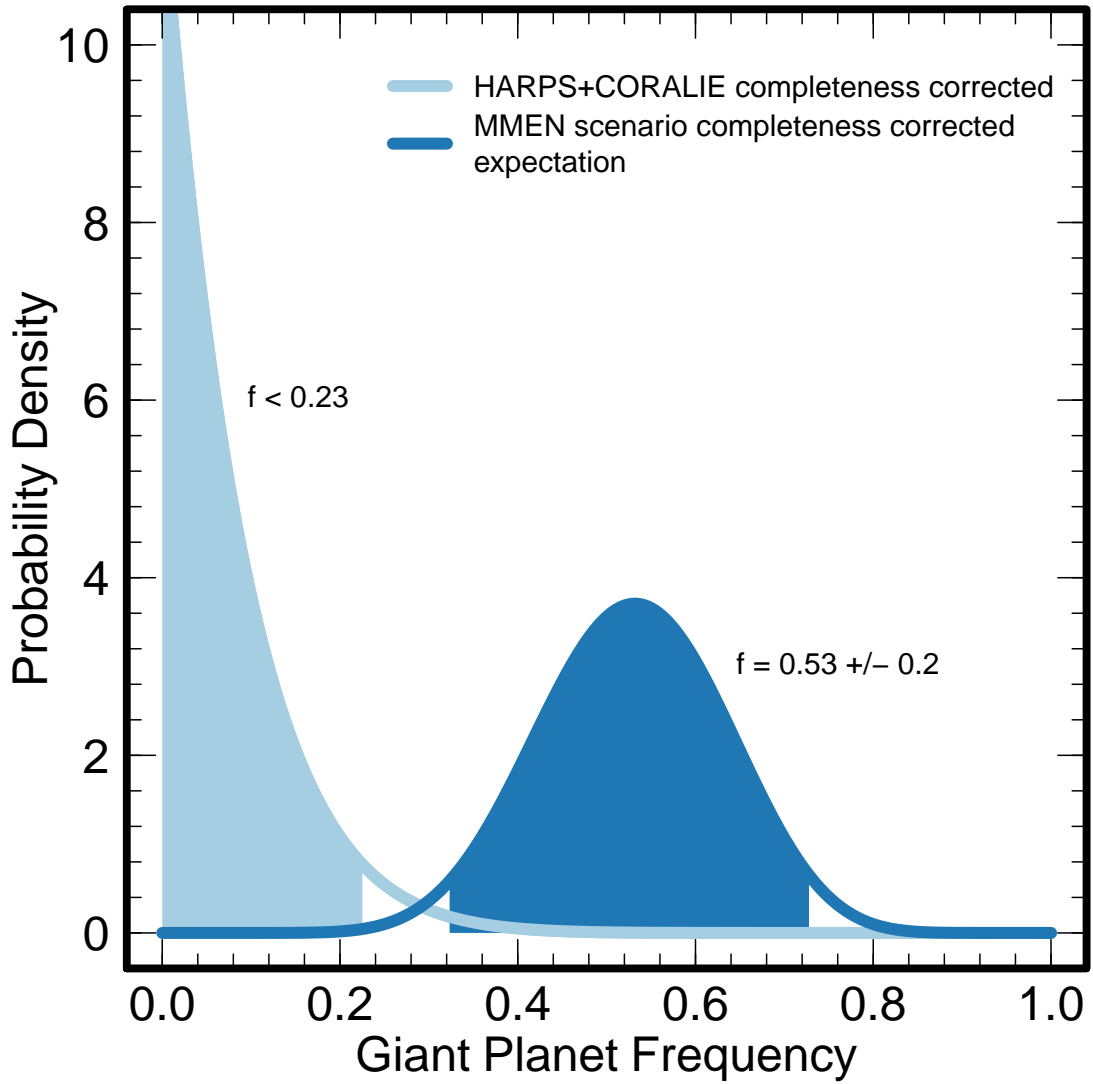}
\caption{Posterior distribution of giant planet occurrence rate.  The
posterior distributions are given by the curves, while the 95\% credible
intervals are indicated by the shaded regions below the curves.  The lack
of announced giant planets in the 20 multiplanet systems discovered
by HARPS combined with the completeness limits given in \citet{may11}
indicate that the upper bound on the 95\% credible interval on giant
planet occurrence in those systems is 0.23.  In the MMEN scenario, giant
planets should be ubiquitous in these systems.  The same completeness
estimates indicate that the observed occurrence rate should be in the
95\% credible interval $0.53 \pm 0.20$.  The probability that the two
distributions overlap is less than one in 1000.\label{fig06}}
\end{figure}

\noindent
I plot these posterior distributions and 95\% credible intervals in
Figure~\ref{fig06}.  I find that the probability that these posterior
distributions overlap to be $p \approx 1 \times 10^{-3}$.  This implies
a 3$\sigma$ (one-sided) difference between the two distributions: there
are too few long-period giant planets observed in RV multiplanet systems
for them to have formed in the MMEN scenario.  The concentration of solid
material in the inner region of the disk hypothesized by \citet{han12}
should make long-period giant planets less common in the RV multiplanet
systems than in the field planet host population.  About 7\% of field FGK
stars have a giant planet with $P$ in the range $1~\mathrm{year} \leq P
\leq 10$ years \citep[e.g.,][]{cum08}.  Unfortunately, in RV multiplanet
systems the current 1$\sigma$ upper limit on long-period giant planet
occurrence is 15\%, so the \citet{han12} scenario is unconstrained.
In the future, a search for long-period giant planets in $n \approx 50$
RV multiplanet systems would be able to resolve the issue.

\subsection{Metallicity Constraints}

The frequency of giant planets can also be constrained by examining the
metallicity distribution of a sample of stars.  Giant planets at all
orbital periods preferentially occur around metal-rich stars, so {\it
Kepler} multiplanet systems that have a long-period giant planet should
preferentially orbit metal-enriched stars.  No strong metallicity effect
has been measured for low-mass or small-radius planets orbiting solar-type
stars, at least for the single systems that dominate the {\it Kepler}
sample \citep[e.g.,][]{sch11,buc12}.

\begin{figure}
\plotone{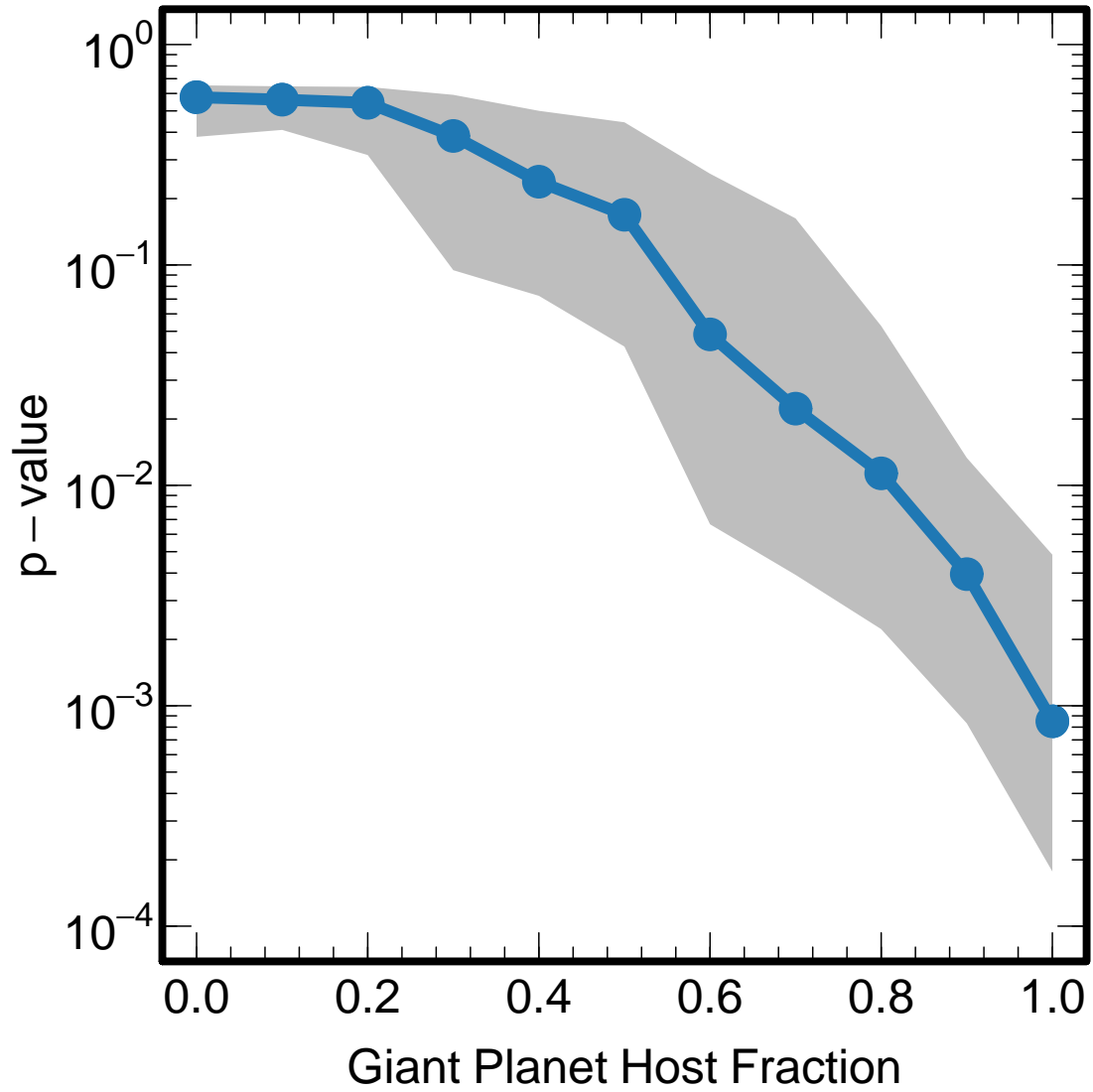}
\caption{Anderson--Darling $p$-value as a function of the fraction of
unobserved giant planet hosts in the sample of {\it Kepler} multiple
small-planet hosts.  The solid blue curve shows the median $p$-value
after bootstrap resampling and the gray polygon shows the 1$\sigma$
confidence interval.  A giant planet host fraction of 1 suggested by the
MMEN scenario is rejected at the $p = 1 \times 10^{-3}$ level, or about
3$\sigma$.
\label{fig07}}

\end{figure}

To quantify the maximum fraction of {\it Kepler} multiplanet systems
that could host a long-period giant planet yet not cause a noticeable
metallicity enhancement in the {\it Kepler} multiplanet sample,
I use a Monte Carlo simulation.  I create many random mixed control
samples composed of the metallicities of both multiple small-planet and
large-planet host stars from \citet{buc12}.  I vary the fraction of large
planet-host stars and then compare the metallicity distribution of the
pure multiple small-planet host sample to the resultant mixed control
sample distributions using the Anderson--Darling Test.  I plot the results
in Figure~\ref{fig07} and find that control samples including less than
$\approx\!\!50\%$ large-planet host stars are generally consistent
with the multiple small-planet host sample.  Consequently, I expect
that no more than 50\% of the multiple small-planet host stars in the
{\it Kepler} field also possess an unobserved long-period giant planet
based on metallicity alone.  This is in contrast to the MMEN expectation,
where nearly all systems of small planets should also have successfully
formed a long-period giant planet companion.  In fact, a giant planet
host fraction of 1 suggested by the MMEN scenario is rejected at the $p =
1 \times 10^{-3}$ level, or about 3$\sigma$ (one-sided).

\section{Discussion}

The MMEN scenario failed two independent tests of its apparent prediction
that long-period giant planets should be ubiquitous in close-in
multiplanet systems, each at $3\sigma$.  This tension indicates that
massive disks alone cannot fully explain the properties of the {\it
Kepler} multiplanet systems.  If in situ formation and migration are
the only two relevant processes, then migration must still have played
a role.  \citet{ray14} reached a similar conclusion after examining disk
surface density profiles.  Therefore, the properties of the {\it Kepler}
multiplanet systems can in principle be used to help determine the rate,
direction, and stopping mechanism for migration in a gaseous disk as
well as their dependencies on planet mass, stellar magnetic field, disk
thermodynamics, turbulence, accretion rate, and dissipation mechanism.

It is possible that increasing the solid surface density of a disk by
increasing metallicity does not have the same effect on giant planet
formation as increasing the solid surface density through overall
disk mass.  It may be that the increased dust mass in a metal-rich
protoplanetary disk efficiently removes ions and therefore affects the
disk structure.  The MRI-inactive regions of a metal-rich disk may
therefore be larger than a solar-metallicity disk, and consequently
promote especially efficient giant planet formation.  In that case,
the increased incidence of giant planets around metal-rich stars may
not be a useful guide to giant planet formation in massive disks.

This study of in situ planet formation has produced many insights that
are useful in the estimation of the frequency of exoplanet systems with
``Solar-System-like" architectures.  Though this was not the aim of this
effort, the topic is of considerably current interest and worth examining
in detail.  Indeed, the presence of long-period giant planets in systems
of terrestrial planets may play an important role in the habitability
of those planets.  \citet{wet94} argued that the formation of Jupiter
prevented the formation of many comets and that Jupiter's subsequent
presence ejected many more comets from the Solar System.  In systems with
no long-period giant planets, the cometary impact flux in the terrestrial
planet region is expected to be 1000 times the observed value in the
Solar System.  Frequent comet impacts may sterilize a planet, and will
definitely inhibit the evolution of intelligent life.  As a result,
the presence of a long-period giant planet in an exoplanet system may
play an important role in its habitability.

The fraction of solar-type stars that host a stable exoplanet system with
at least one low-mass planet and at least one long-period giant planet is

\begin{eqnarray}
\eta_{ss} = \eta_{\mathrm{single}}\,\eta_{\mathrm{sp}}\,
            \eta_{\mathrm{gp}}\,\eta_{\mathrm{Hill}},
\end{eqnarray}

\noindent
where $\eta_{\mathrm{single}}$ is the fraction of solar-type stars
in single star systems, $\eta_{\mathrm{sp}}$ is the fraction of
solar-type stars that host small planets, $\eta_{\mathrm{gp}}$
is the fraction of solar-type stars that host a giant planet, and
$\eta_{\mathrm{Hill}}$ is the fraction of systems that are Hill stable.
The factor $\eta_{\mathrm{Hill}}$ is necessary because I assume that
$\eta_{\mathrm{sp}}$ and $\eta_{\mathrm{gp}}$ are independent of each
other and planet period.  This assumption would lead to an overestimate of
$\eta_{ss}$, as some of the hypothetical systems would be unstable.  As a
result, I correct after-the-fact using the factor $\eta_{\mathrm{Hill}}$
to ensure that I only count systems that would not be obviously unstable.

About 1/3 of the FGK stars in the solar neighborhood are in single
systems, so $\eta_{\mathrm{single}} = 0.33$ \citep[e.g.,][]{duq91}.
\citet{fre13} report that 40\% of solar-type stars have at least one
planet with $P < 50$ days, so $\eta_{\mathrm{sp}} = 0.4$.  The factor
$\eta_{\mathrm{gp}}$ depends sensitively on metallicity, so it is
a function of the metallicity distribution of a stellar population.
Because giant planet occurrence is such a strong function of metallicity,
Galactic giant planets are only likely to be found in the Milky Way's disk
or bulge.  For the disk metallicity distribution, I use the distribution
of [M/H] from \citet{cas11}.  For the bulge metallicity distribution,
I use the distribution of [M/H] from \citet{ben13}, assuming that the
abundances inferred from microlensed dwarf stars studied in that survey
are representative of the bulge metallicity distribution.  Also for the
bulge sample, non-solar abundance patterns are important to consider.  For
that reason, I compute [M/H] from the measured values of [Fe/H], [O/Fe],
[Mg/Fe], and [Si/Fe] assuming the Solar abundances from \citet{asp05}.
These four elements contribute 99\% of the total stellar metallicity of
the available abundances in the \citet{ben13} catalog, and I therefore
neglect the other elements.  To calculate $\eta_{\mathrm{gp}}$ for each
population, I use

\begin{eqnarray}
\eta_{\mathrm{gp}} = \frac{1}{m} \sum_{i=1}^{m} P_{\mathrm{gp}}(\mathrm{[M/H]}_i),
\end{eqnarray}

\noindent
where $m$ is the number of measured metallicities in a catalog
and $P_{\mathrm{gp}}(\mathrm{[M/H]_i})$ is Equation (7).  I find that
$\eta_{\mathrm{gp}} = 0.06$ for the disk and $\eta_{\mathrm{gp}} = 0.17$
for the bulge.

Plugging in the numbers, I find that $\eta_{ss} = 0.0067$ in the disk and
$\eta_{ss} = 0.017$ in the bulge.  Recently, \citet{pet13} published
a tentative extrapolation of the frequency of Earth-size planets with
orbital periods in the range $200~\mathrm{days} \leq P \leq 400$ days;
they found that $\eta_{\mathrm{sp}} = \eta_{\oplus} \approx 0.06$.
An Earth-mass planet orbiting at $a = 1$ AU implies that any giant planets
in the system must be beyond $a \approx 3.5$ AU to avoid Hill instability.
Using the giant planet orbital distribution from \citet{cum08}, that
implies that $\eta_{\mathrm{Hill}} = 0.44$.  As a result, the fraction
of solar-type stars hosting true Solar System analogs including both
an Earth-size planet near 1 AU and a long-period Jupiter-mass planet is
$\eta_{ss} = 0.00059$ in the disk and $\eta_{ss} = 0.0015$ in the bulge.

The total number of such system in the Galaxy is

\begin{eqnarray}
N_{ss} = \eta_{\mathrm{FGK}}\,
         \left(\eta_{ss,\mathrm{disk}}\,M_{\mathrm{disk}}+
               \eta_{ss,\mathrm{bulge}}\,M_{\mathrm{bulge}}\right),
\end{eqnarray}

\noindent
where $\eta_{\mathrm{FGK}}$ is the FGK mass fraction in a stellar
population, $M_{\mathrm{disk}}$ is the stellar mass of the Milky
Way's disk, and $M_{\mathrm{bulge}}$ is the stellar mass of the
Milky Way's bulge.  The stellar mass of the Milky Way's disk is
$M_{\mathrm{disk}} \approx 4.6 \times 10^{10} M_{\odot}$, while the
stellar mass of the bulge is $M_{\mathrm{bulge}} \sim 1 \times 10^{10}
M_{\odot}$ \citep{bov13,wid08}.  I assume a \citet{cha03} initial mass
function (IMF) truncated at $0.08\,M_{\odot}$ and $120\,M_{\odot}$ for
both the disk and bulge population, as there is no conclusive evidence
that the IMF varies between the two populations \citep[e.g.,][]{bas10}.
Given the \citet{cha03} IMF, I use the algorithm from the \texttt{SLUGS}
code\footnote{
   https://sites.google.com/site/runslug/ and described in \citet{fum11}
   and \citet{das12}.} to randomly sample the IMF and compute the fraction
of stellar mass in FGK stars ($0.8\,M_{\odot} \leq M_{\ast} \leq
1.25\,M_{\odot}$).  I find that 10\% of a stellar population's stellar
mass is in FGK stars, and because each star has $M_{\ast} \approx
1\,M_{\odot}$, the number of stars is equivalent to the stellar mass in
FGK stars.  As a result, $\eta_{\mathrm{FGK}} = 0.1$.

Putting all of the factors together, I find that the number of ``solar
systems" in the Galaxy with at least one Neptune-size planet with
orbital period $P < 50$ days and a long-period giant planet is $N_{ss}
\sim 5 \times 10^{7}$.  The number of true Solar System analogs in the
Galaxy with an Earth-size planet in the habitable zone and a long-period
Jupiter-mass giant planet is $N_{ss} \sim 4 \times 10^{6}$.

In this analysis, I have not accounted for the possibility of metallicity
gradients in the disk of the Milky Way.  The inner disk has a higher
stellar density than the solar neighborhood, and it is likely more
metal rich too \citep[e.g.,][]{fri13}.  Similarly, the outer disk has a
lower stellar density than the solar neighborhood and may be metal poor
as well.  These issues are both very active areas of a research and
future data from the APO Galactic Evolution Experiment (APOGEE) or the
Gaia-ESO Survey may resolve the issue.

\section{Conclusions}

{\it Kepler} multiple planet systems with several planets orbiting with
periods $P < 50$ days may be difficult to explain in the traditional
core accretion and Type I migration paradigm.  In response, two in situ
models of plant formation were proposed: the minimum mass extrasolar
nebula (MMEN) scenario of \citet{chi13} and the planetesimal migration
scenario of \citet{han12}.  Both models make predictions for the
occurrence rate of long-period giant planets.  In the MMEN scenario,
they are ubiquitous.  In the planetesimal migration scenario, they are
less common than in the field population.  I find that the prediction
from the MMEN scenario for the occurrence of long-period giant planets in
multiple low-mass systems discovered with the radial velocity technique
fails at 3$\sigma$.  The lack of metallicity enhancement in the hosts
of multiple small-planet systems discovered by {\it Kepler} provides an
independent test of the MMEN scenario, which it also fails at 3$\sigma$.
I am unable to constrain the planetesimal accretion scenario with the
current sample.  As a result, migration is still a necessary step in
the formation of systems of close-in low-mass planets.

I also rederived the scaling of giant planet occurrence on metallicity,
and I find that $P_{\mathrm{gp}} = 0.05_{-0.02}^{+0.02} \times 10^{(2.1
\pm 0.4) [\mathrm{M/H}]} = 0.08_{-0.03}^{+0.02} \times 10^{(2.3 \pm 0.4)
[\mathrm{Fe/H}]}$.  I used these relations to calculate the frequency
of ``solar systems" in the Galaxy, where a solar system is defined
as a single FGK star orbited by a planetary system with at least one
small planet interior to the orbit of a giant planet.  The presence
of a giant planet exterior to an Earth-size planet may be necessary to
prevent frequent comet impacts from inhibiting the evolution of life on
an otherwise habitable planet.  I find that in the solar neighborhood,
about 0.7\% of solar-type stars have a ``solar system" consisting of both
a Neptune-size planet with $P < 50$ days and a protective long-period
companion.  Intriguingly, I find that true Solar System analogs with both
a terrestrial planet in the habitable zone and a long-period giant planet
companion to protect it occur around only 0.06\% of solar-type stars.
There are perhaps $4 \times 10^{6}$ such systems in the Galaxy.

\acknowledgments I thank Kat Deck, Greg Laughlin, Roberto Sanchis-Ojeda,
and Josh Winn.  This research has made use of NASA's Astrophysics Data
System Bibliographic Services, the Exoplanet Orbit Database and the
Exoplanet Data Explorer at exoplanets.org, the Exoplanet Encyclopedia at
exoplanets.eu, and both the SIMBAD database and VizieR catalogue access
tool, CDS, Strasbourg, France.  The original description of the VizieR
service was published by \citet{och00}.  Some of the data presented in
this paper were obtained from the Mikulski Archive for Space Telescopes
(MAST).  STScI is operated by the Association of Universities for Research
in Astronomy, Inc., under NASA contract NAS5-26555. Support for MAST
for non-HST data is provided by the NASA Office of Space Science via
grant NNX13AC07G and by other grants and contracts.  Support for this
work was provided by the MIT Kavli Institute for Astrophysics and Space
Research through a Kavli Postdoctoral Fellowship.

\clearpage

\clearpage

\clearpage

\clearpage

\clearpage

\clearpage

\clearpage

\clearpage
\begin{deluxetable}{rrrrrrr}
\tablecaption{Multiple Low-Mass Planet Systems\label{tbl-1}}
\tablewidth{0pt}
\tablehead{\colhead{Name} & \colhead{System} & \colhead{$P$} & \colhead{$e$} &
\colhead{$K$} & \colhead{$m\sin{i}$} & \colhead{Reference} \\
\colhead{} & \colhead{} & \colhead{[days]} & \colhead{} &
\colhead{[m s$^{-1}$]} & \colhead{[$M_{\oplus}$]} & \colhead{}}
\startdata
HD 1461 b       & HD 1461       & 5.773 & 0.14  & 2.44  & 7.6   & \citet{riv10} \\
HD 1461 c       & HD 1461       & 13.5  & 0     & 1.57  & 5.9   & \citet{may11} \\
HD 10180 c      & HD 10180      & 5.76  & 0.08  & 4.54  & 13    & \citet{lov11} \\
HD 10180 d      & HD 10180      & 16.36 & 0.14  & 2.93  & 12    & \citet{lov11} \\
HD 10180 e      & HD 10180      & 49.75 & 0.06  & 4.25  & 25    & \citet{lov11} \\
HD 10180 f      & HD 10180      & 122.7 & 0.13  & 2.95  & 24    & \citet{lov11} \\
HD 10180 g      & HD 10180      & 602   & 0     & 1.56  & 21    & \citet{lov11} \\
HD 10180 h      & HD 10180      & 2248  & 0.15  & 3.11  & 66    & \citet{lov11} \\
HD 13808 b      & HD 13808      & 14.18 & 0.17  & 3.53  & 10    & \citet{may11} \\
HD 13808 c      & HD 13808      & 53.83 & 0.43  & 2.81  & 11    & \citet{may11} \\
HD 20003 b      & HD 20003      & 11.85 & 0.4   & 4.03  & 12    & \citet{may11} \\
HD 20003 c      & HD 20003      & 33.82 & 0.16  & 2.95  & 13    & \citet{may11} \\
HD 20794 b      & HD 20794      & 18.32 & 0     & 0.83  & 2.7   & \citet{pep11} \\
HD 20794 c      & HD 20794      & 40.11 & 0     & 0.56  & 2.4   & \citet{pep11} \\
HD 20794 d      & HD 20794      & 90.31 & 0     & 0.85  & 4.7   & \citet{pep11} \\
HD 20781 b      & HD 20781      & 29.15 & 0.11  & 3.03  & 12    & \citet{may11} \\
HD 20781 c      & HD 20781      & 85.13 & 0.28  & 2.88  & 16    & \citet{may11} \\
HD 21693 b      & HD 21693      & 22.66 & 0.26  & 2.73  & 10    & \citet{may11} \\
HD 21693 c      & HD 21693      & 53.88 & 0.24  & 4.02  & 21    & \citet{may11} \\
HD 31527 b      & HD 31527      & 16.55 & 0.13  & 3.01  & 12    & \citet{may11} \\
HD 31527 c      & HD 31527      & 51.28 & 0.11  & 2.83  & 16    & \citet{may11} \\
HD 31527 d      & HD 31527      & 274.5 & 0.38  & 1.79  & 17    & \citet{may11} \\
HD 39194 b      & HD 39194      & 5.636 & 0.2   & 1.95  & 3.7   & \citet{may11} \\
HD 39194 c      & HD 39194      & 14.02 & 0.11  & 2.26  & 5.9   & \citet{may11} \\
HD 39194 d      & HD 39194      & 33.94 & 0.2   & 1.49  & 5.2   & \citet{may11} \\
HD 40307 b      & HD 40307      & 4.312 & 0     & 1.97  & 4.1   & \citet{may09} \\
HD 40307 c      & HD 40307      & 9.62  & 0     & 2.47  & 6.7   & \citet{may09} \\
HD 40307 d      & HD 40307      & 20.46 & 0     & 2.55  & 8.9   & \citet{may09} \\
HD 51608 b      & HD 51608      & 14.07 & 0.15  & 4.10  & 13    & \citet{may11} \\
HD 51608 c      & HD 51608      & 95.42 & 0.41  & 3.25  & 18    & \citet{may11} \\
HD 69830 b      & HD 69830      & 8.667 & 0.1   & 3.51  & 10    & \citet{lov06} \\
HD 69830 c      & HD 69830      & 31.56 & 0.13  & 2.66  & 12    & \citet{lov06} \\
HD 69830 d      & HD 69830      & 197   & 0.07  & 2.2   & 18    & \citet{lov06} \\
HD 93385 b      & HD 93385      & 13.19 & 0.15  & 2.21  & 8.4   & \citet{may11} \\
HD 93385 c      & HD 93385      & 46.02 & 0.24  & 1.82  & 10    & \citet{may11} \\
HD 96700 b      & HD 96700      & 8.126 & 0.1   & 3.02  & 9     & \citet{may11} \\
HD 96700 c      & HD 96700      & 103.5 & 0.37  & 1.98  & 13    & \citet{may11} \\
HD 109271 b     & HD 109271     & 7.854 & 0.25  & 5.6   & 17    & \citet{lo13} \\
HD 109271 c     & HD 109271     & 30.93 & 0.15  & 4.9   & 24    & \citet{lo13} \\
61 Vir b        & 61 Vir        & 4.215 & 0.12  & 2.12  & 5.1   & \citet{vog10} \\
61 Vir c        & 61 Vir        & 38.02 & 0.14  & 2.12  & 11    & \citet{vog10} \\
61 Vir d        & 61 Vir        & 123   & 0.35  & 3.25  & 23    & \citet{vog10} \\
HD 134606 b     & HD 134606     & 12.08 & 0.15  & 2.68  & 9.3   & \citet{may11} \\
HD 134606 c     & HD 134606     & 59.52 & 0.29  & 2.17  & 12    & \citet{may11} \\
HD 134606 d     & HD 134606     & 459.3 & 0.46  & 3.66  & 38    & \citet{may11} \\
HD 136352 b     & HD 136352     & 11.58 & 0.18  & 1.77  & 5.3   & \citet{may11} \\
HD 136352 c     & HD 136352     & 27.58 & 0.16  & 2.82  & 11    & \citet{may11} \\
HD 136352 d     & HD 136352     & 106.7 & 0.43  & 1.68  & 9.5   & \citet{may11} \\
HD 192310 b     & HD 192310     & 74.72 & 0.13  & 3     & 17    & \citet{pep11} \\
HD 192310 c     & HD 192310     & 525.8 & 0.32  & 2.27  & 23    & \citet{pep11} \\
HD 215152 b     & HD 215152     & 7.282 & 0.34  & 1.26  & 2.8   & \citet{may11} \\
HD 215152 c     & HD 215152     & 10.87 & 0.38  & 1.26  & 3.1   & \citet{may11}
\enddata
\end{deluxetable}

\clearpage
\begin{deluxetable}{rrrrrr}
\tablecaption{Host Stars of Multiple Low-Mass Planet Systems\label{tbl-2}}
\tablewidth{0pt}
\tablehead{\colhead{Name} & \colhead{HIP} & \colhead{HD} & \colhead{$M_{V}$} &
\colhead{$B-V$} & \colhead{$N_{p}$} \\
\colhead{} & \colhead{} & \colhead{} & \colhead{[mag]} &
\colhead{[mag]} & \colhead{}} 
\startdata
HD 1461         &   1499        &   1461        & 4.63  & 0.67  &  2 \\
HD 10180        &   7599        &  10180        & 4.37  & 0.63  &  6 \\
HD 13808        &  10301        &  13808        & 6.10  & 0.87  &  2 \\
HD 20003        &  14530        &  20003        & 5.17  & 0.77  &  2 \\
HD 20794        &  15510        &  20794        & 5.36  & 0.71  &  3 \\
HD 20781        &  15526        &  20781        & 5.71  & 0.82  &  2 \\
HD 21693        &  16085        &  21693        & 5.40  & 0.76  &  2 \\
HD 31527        &  22905        &  31527        & 4.56  & 0.61  &  3 \\
HD 39194        &  27080        &  39194        & 6.02  & 0.76  &  3 \\
HD 40307        &  27887        &  40307        & 6.59  & 0.94  &  3 \\
HD 51608        &  33229        &  51608        & 5.46  & 0.77  &  2 \\
HD 69830        &  40693        &  69830        & 5.47  & 0.75  &  3 \\
HD 93385        &  52676        &  93385        & 4.37  & 0.60  &  2 \\
HD 96700        &  54400        &  96700        & 4.47  & 0.61  &  2 \\
HD 109271       &  61300        & 109271        & 4.11  & 0.66  &  2 \\
61 Vir          &  64924        & 115617        & 5.08  & 0.71  &  3 \\
HD 134606       &  74653        & 134606        & 4.75  & 0.74  &  3 \\
HD 136352       &  75181        & 136352        & 4.80  & 0.64  &  3 \\
HD 192310       &  99825        & 192310        & 5.98  & 0.88  &  2 \\
HD 215152       & 112190        & 215152        & 6.47  & 0.97  &  2
\enddata
\end{deluxetable}

\appendix
\section{Kepler Multiplanet System Query}

The following SQL query can be used in the {\it Keper} CasJobs database
available from MAST to reproduce my sample of {\it Kepler} multiple
small-planet candidate systems.  It is first necessary to set ``Context"
to ``kepler".

\begin{verbatim}
SELECT a.kepid, a.kepoi_name, a.koi_prad, a.koi_period, a.koi_srad,
c.kepler_name
FROM kepler_koi a
INNER JOIN (
   SELECT kepid
   FROM kepler_koi
   WHERE koi_disposition != 'FALSE POSITIVE'
   AND koi_prad < 5
   GROUP BY kepid
   HAVING COUNT(kepoi_name) > 1
) b ON a.kepid = b.kepid
LEFT OUTER JOIN published_planets c
ON a.kepid = c.kepid AND a.kepoi_name = c.kepoi_name
INNER JOIN keplerObjectSearchWithColors d ON a.kepid = d.kic_kepler_id
WHERE a.koi_disposition != 'FALSE POSITIVE'
AND d.jh BETWEEN 0.22 AND 0.62
AND d.hk BETWEEN 0.00 AND 0.10
AND d.jk BETWEEN 0.22 AND 0.72
ORDER BY a.kepoi_name, a.koi_period;
\end{verbatim}

\end{document}